\documentclass{article}
\usepackage{graphicx}
\usepackage{color}
\usepackage{cite}

\def\rRichClub{\Phi}

\def\text#1{\rm{#1}}

\def\linksRC{k^+}

\def \probRankRank{P}

\def\degree{k}

\def\mapCons{{f}}
\def\qProb{{q}}

\def\lang{{\lambda}}
\def\avDegDeg{{k_{\rm nn}}}
\def\eProb{{p}}
\def\massFunct{{u}}
\def\stepFunct{{w}}
\def\entrop{S}
\def\highDeg{k^+}

\begin{document}
\title{Network Null Model based on Maximal Entropy and the Rich--Club}
\author{ 
R. J. Mondrag\'on\\
Queen Mary University of London, \\
School of Electronic Engineering and Computer Science, \\
Mile End Road, London, E1 4NS, UK\\
r.j.mondragon@eecs.qmul.ac.uk
}

\maketitle


\begin{abstract}
{We present a method to construct a network null--model based on the maximum entropy principle and where the restrictions that the rich--club and the degree sequence impose are conserved. We show that the probability that two nodes share a link can be described with a simple  probability function.  The null--model closely approximates the assortative properties of the network.}
{}
\end{abstract}

\maketitle

\section{Introduction}
In complex networks there are topological patterns that are presumed to be important to  network structure and its behaviour. The question is how to validate if patterns are a random occurrence or if there is a mechanism, perhaps unknown, responsible for them. A common technique in analysing properties of a complex network is to use a statistical randomisation method to create an ensemble of surrogate networks. The ensemble is used as a reference null--model for comparison purposes. To take into consideration the intrinsic structure of the network, the randomisation procedures are restricted to preserve the number of connections of individual nodes, the degree sequence, as this is considered a basic property of a network~\cite{Newman01}. Two common procedures to generate surrogates with a given degree sequence are the reshuffling of end points of a pair of links ~\cite{Molloy98,Maslov02}, or to creating ``stubs'' nodes with the desired degree and then joining a pair of randomly selected stubs to form a link~\cite{Doro03}. Both procedures have disadvantages, the reshuffling procedure can create sampling biases~\cite{Maslov02,Maslov04,farkas2004equilibrium,Maslov03Book}
and the stubs procedure can create undesired networks, for example, self loops~\cite{Newman01}. Many of these difficulties can be overcome using sophisticated statistical methods, for example, to remove the bias in the reshuffling method~\cite{annibale2009tailored,coolen2011modelling}.
%
An extension in the generation of null--models is to take into consideration network properties beyond the degree distribution. The aim is to generate random networks with \emph{a priori} degree distribution and degree--degree correlations. A common technique to generate this random network is to extend the stub method to include higher order statistics~\cite{weber2007generation,raschke2010generation}. The null--model is generated by averaging over an ensemble of networks generated by these methods.

Shannon's entropy and related information measures are widely in use to describe the complexity and properties of networks~\cite{gfeller2005finding,bogacz2006homogeneous,gomez2008entropy,Bianconi08a,Anand09,Dehmer11,Bianconi09,bauer2002maximal,sinatra2011maximal,wang2006entropy}.
For instance, the entropy of the degree distribution has been used to obtain networks that are robust to networks attacks~\cite{wang2006entropy}. 
%
Shannon's entropy can also be used to obtain a null--model that best describes our state of knowledge of the network structure. In this case the maximal entropy approach (MaxEnt) is used to describe our state of knowledge in a way that is ``maximally noncommittal'' by certain criterion~\cite{Jaynes57}. The maximal entropy method is attractive because it produces null--models with probabilistic characteristics only warranted by the data. 
The maximisation of entropic measures have been used to describe the complexity of networks~\cite{Bianconi08a,Anand09,Dehmer11} and to construct null--models that conserve the degree distribution $P(k)$, the degree--degree correlation $P(k_i,k_j)$ and the community structure~\cite{Bianconi08a,Bianconi09}.  
%
%
The MaxEnt approach has many other uses. It has been used to design random walks in a network such that the random walkers explore every possible walk with equal probability~\cite{bauer2002maximal,sinatra2011maximal}. 
It has also been used 
to propose that  scale--free networks defined only by their degree sequence, are most likely to be disassortative~\cite{Johnson10}. This result was obtained using the ansatz that the average degree of the nearest neighbour of a node can be described with a power law. 

As a method to construct a null--model the MaxEnt method has been used to  measure how a structural constraint, like the degree sequence $P(k)$ or/and the degree--degree correlations $P(k',k)$,  shape the network structure~\cite{Bianconi09}.
A common difficulty shared by the randomisation, stubs and MaxEnt methods in the creation of null--models based on the conservation of the degree sequence and degree--degree correlation is the lack of a faithful portray of the degree--degree correlation.
In a network with heavy--tailed degree distribution it is not possible to obtain from measurements a reliable characterisation of the degree--degree correlation $P(k',k)$, in particular for large degrees, due to insufficient data~\cite{Pastor01,Doro03}. To overcome this limitation it is common to use the average degree of the nearest neighbour of a node with degree $k$, $\langle \avDegDeg (k)\rangle =\sum_{k'} k'P(k'|k)$. This quantity is use to classify the assortativiness of a network~\cite{Pastor01}. This quantity has some limitations: it conveys less information than the degree--degree correlation, it is the average of an average, and could be ambiguous when classifying the assortativity of a network~\cite{dorogovtsev2010lectures}.  These limitations mean that using $P(k',k)$ or $\langle \avDegDeg (k)\rangle$ as network constraints generate statistical inaccuracies in the null--model 
which will affect our ability to discern if a network property could have been caused by chance. 

In here we present a method to construct a MaxEnt null--model that is defined by the degree sequence and the rich--club coefficient~\cite{Zhou2004}. 
{
The rich--club coefficient measures the density of connections between nodes with degree higher or equal to a given degree. 
We conserve the rich--club since it is known that it is a projection of the degree--degree correlation~\cite{Colizza06,Xu10} and can be evaluated accurately from the data. 
}

\section{The method: Rich--club and maximal entropy}
{ To distinguish the nodes, we  rank them
}
 in decreasing order of their  degrees,
{
the node with the highest degree is ranked first and so on.
 }  As a consequence of the ranking, two nodes with equal degrees will be distinguishable as they would have different ranks. Under this ranking scheme we also evaluate the number of links $\highDeg_r$ that  node $r$ shares with nodes of higher rank, $r'\le r$. 
The number of links that a node has is divided into the number of links with nodes of higher rank
$\highDeg_r$ and the number of links with nodes of lower rank $\degree_r-\highDeg_r$.
Notice that $\highDeg_r$ could be large enough to allow multiple links between two nodes.
The network is characterised by the sequences  $\{ k_1, k_2 \ldots k_N\}$ and $\{\highDeg_1, \highDeg_2 \ldots \highDeg_N\}$. 
The total number of links in the network  is $L=\sum_{i=1}^N k_i$  
and the rich--club coefficient~\cite{Zhou2004} is $\rRichClub_r= 2\sum_{i=1}^{r}  \linksRC_i /(r(r-1))$. An ensemble of networks that have the same sequence  $\{\highDeg_1, \highDeg_2 \ldots \highDeg_N\}$ will also have the same  rich--club $\rRichClub_r$.

\begin{figure}
 \begin{center}
 \includegraphics[width=8cm]{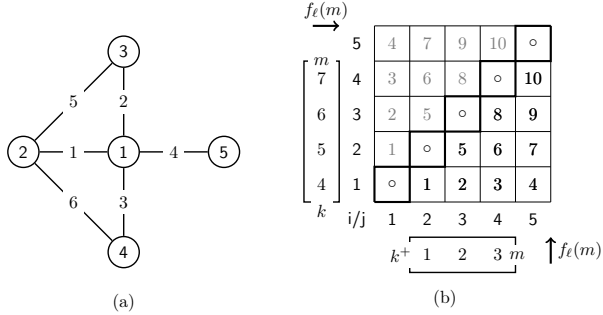}
 \end{center}
 \caption{\label{fig:maps} { (a) A five node network where the nodes are labelled in decreasing order of their degree and the links' labels are given by the map $\ell=g(i,j)$. (b)} Relationship between the links' label (inside the matrix), with the nodes' rank ({$i/j$}) and the constraint equations labels $m$ (show inside the rectangles).  As an example how to read this table, the constraint equation $m=3$ is related to conservation of $\linksRC_4$ which is given by the sum of the probabilities of links 3, 6 and 8. These links correspond to the connections between the pair of nodes 1--4, 2--4 and 3--4. The constraint equation $m=5$ is related to the conservation of $k_2$ which is given by the sum of the probabilities of links 1, 5, 6, and 7.}
 \end{figure}

 Let us assume that $\probRankRank_{r',r}$ is the probability that node $r$ connects to node $r'$ and that $\probRankRank_{r,r}=0$ as self--loops are not allowed. Given  the $\linksRC_r$ links,  we  constrain the connectivity of a network by imposing the condition that the 
 {
 expected value of the 
}
 number of links, $\langle{\linksRC_r}\rangle$ satisfies
\begin{equation}
\label{eq:probRestriction}
\langle {\linksRC_r} \rangle= \sum_{i=1}^{r-1} \probRankRank_{i,r}=\linksRC_r,
\end{equation}
and the 
 {
 expected value of the
}
 degree $\langle k_r \rangle$ is
\begin{equation}
\label{eq:estimation-k} 
\langle k_r \rangle=\sum_{j=1}^N\probRankRank_{j,r}={\linksRC_r} + \sum_{j=r+1}^N \probRankRank_{j,r}=k_r.
\end{equation} 
{
To formulate the MaxEnt problem it is convenient to normalise the above constraints, that is to consider $\{k_1/L,\ldots ,k_N/L \}$ and $\{\linksRC_1/L,\ldots ,\linksRC_N/L \}$
 instead of $\{k_1,\ldots ,k_N \}$ and $\{\linksRC_1,\ldots ,\linksRC_N \}$, and to represent the interaction between node $i$ and node $j$ with the link $\ell$ that joins them.  The MaxEnt solution is formulated using the probabilities $\eProb_\ell$ of the normalised constraints instead of $P_{i,j}$.
}
 If the interaction between nodes $i$ and  $j$ is represented by a random variable
 and 
 if this interaction is labeled by $\ell=g(i,j)=(i-1)N-i(1+i)/2+j$
 if $i>j$, then the
  entropy associated with this interaction is 
 $s(\eProb_\ell)=-\eProb_\ell \log \eProb_\ell$ 
where  $\eProb_\ell = \eProb_{g(i,j)}$ is the probability that $i$ and $j$ interact via a link.
 The total entropy of the network is
%
 \begin{equation}
 \entrop(\eProb_1, \ldots, \eProb_{N(N-1)/2}) = -\sum_{\ell=1}^{N(N-1)/2} \eProb_\ell \log \eProb_\ell. 
 \end{equation}
{ 
The maximal entropy solution is  the set of  probabilities $\eProb_\ell$ where the entropy $S$ is maximal under certain constraints.
In here,  the constraints are}
%
the normalisation
{
\begin{equation}
\label{eq:norma}
\sum_{\ell=1}^{N(N-1)/2} \eProb_\ell=1
\end{equation}
the conservation of $\linksRC_r$
\begin{equation}
\label{eq:constraintOne}
 \sum_{i=1}^{r-1} p_{g(i,r)}=\frac{\linksRC_r}{L},\quad r=1,\ldots, N-2
\end{equation}
and the conservation of $k_r$
\begin{equation}
\label{eq:constraintTwo}
\sum_{j=1}^N p_{g(j,r)}=\frac{\linksRC_r}{L} + \sum_{j=r+1}^N p_{g(j,r)}=\frac{k_r}{L},\quad r=1, \ldots, N-1.
\end{equation} 
The common procedure to obtain the MaxEnt solution uses the transformation $\eProb_\ell = \exp(-\qProb_\ell)$ and represents the  constraints Eq.~(\ref{eq:constraintOne})--(\ref{eq:constraintTwo}) as the single relationship
\begin{equation}
\label{eq:lagConstr}
\sum_{\ell=1}^{N(N-1)/2} \mapCons_m(\ell)e^{-\qProb_\ell}=c_m,\quad m=1,\ldots, M
\end{equation}
}
where  $c_m$ are $M$ constraints that are related to  $\qProb_\ell$ via the map $\mapCons_m(\ell)$.
For the case of $c_m=k_m/L$ the number of constraints is $N-1$ corresponding to the number of degrees conserved. The number of constraints is $N-1$ and not $N$ as the total number of links $L=\sum_i k_i$ is conserved by the normalisation of $\eProb_\ell$. For the case $c_m=\linksRC_m/L$ the number of constraints is $N-2$ as by construction $\linksRC_1=0$ and $\linksRC_N=k_N$,
giving $M=2N-2$ constraints.
To clarify the relationship between the  links, nodes and constraints labels, Fig.~\ref{fig:maps} shows their relationship for a five nodes network.

If the Lagrangian multipliers are $\lang_0,\ldots\lang_M$ then the MaxEnt solution is obtained by the maximisation of the function
${\cal F}(\qProb_1,\ldots, \qProb_{N(N-1)/2})=\sum_\ell^{N(N-1)/2)}(\qProb_\ell+\lang_0)\,e^{-\qProb_\ell}+\sum_m^M\lang_m\sum_\ell^{N(N-1)/2}f_m(\ell)\,e^{-\qProb_\ell}$. The maximisation condition $\partial {\cal F}((\qProb_1,\ldots, \qProb_{N(N-1)/2})/\partial \qProb_\ell=0$ for $\ell=1,\ldots, N(N-1)/2$ gives the MaxEnt solution 
\begin{equation}
\label{eq:solMaxEnt3}
\qProb_\ell =  1-\lang_0-\sum_{m=1}^M \lang_m f_m(\ell).
\end{equation}
{
Equations~(\ref{eq:norma})--(\ref{eq:constraintTwo}) and (\ref{eq:solMaxEnt3}) define $N(N-1)/2+M+1$ equations with the $N(N-1)/2+M+1$ unknowns $\eProb_1,\ldots,\eProb_{N(N-1)/2}, \lang_0,\ldots,\lang_M$.
%
Usually the solution of the MaxEnt problem is formulated using the Partition function formalism as it gives a smaller set of non--linear equations which are solved numerically. Here we do not use this formalism as we can obtain a recursive solution for the set of equations. The main observation to obtain this recursive solution comes from the term $ f_m(\ell)$ in Eq.~(\ref{eq:solMaxEnt3}).  Given a link $\ell=g(i,j)$ there are only two values of $m$ where $f_m(\ell)$ contributes to  Eq.~(\ref{eq:solMaxEnt3}).
For example in Fig.~\ref{fig:maps}~(b), the link $\ell=6$  is related to the values of $m=3$ (read vertically from the table), and $m=5$ (read horizontally from the table), these two values can be related to the rank $i$ and $j$ labels, in this case $j=4$ gives $m=j-1=3$ and $i=2$ gives $m=N-2+i=5-2+2=5$. 

If the Lagrangian multipliers are labelled using the nodes' rank instead of the label $m$ then Eq.~(\ref{eq:solMaxEnt3}) becomes
}

\begin{equation}
\label{eq:langOne}
\eProb_{g(i,j)} = e^{-1+\lambda_0+\lambda_{j-1}}e^{\lambda_{N-2+i}},
\end{equation}
where we used $\eProb_{g(i,j)} =e^{-\qProb_{g(i,j)}}$. 

From the constraints given by Eq.~(\ref{eq:constraintOne}) these probabilities satisfy 
\begin{equation}
\label{eq:useConstOne}
\sum_{i=1}^{j-1} \eProb_{g(i,j)} = e^{-1+\lambda_0+\lambda_{j-1}}\sum_{i=1}^{j-1}e^{\lambda_{N-2+i}} = \frac{\linksRC_j}{L}. 
\end{equation}
If 
$\massFunct(i) =  e^{\lambda_{N-2+i}}$ 
then the probability function $\eProb_{g(i,j)} = e^{-1+\lambda_0+\lambda_{j-1}}e^{\lambda_{N-2+i}}$
{ 
}
 can be written as
\begin{equation}
\label{eq:massResult}
\eProb_{g(i,j)} = \left(\frac{\massFunct(i)}{\sum_{m=1}^{j-1}\massFunct(m)}\right) \frac{\linksRC_j}{L}, \quad i<j\le N, 
\end{equation}
where 
$e^{-1+\lambda_0+\lambda_{j-1}}=(1/\sum_{i=1}^{j-1}e^{\lambda_{N-2+i}})(\linksRC_j/L)$. 
The probability that there is a link between nodes $i$ and $j$ can be expressed with two factors.
The factor $\linksRC_j/L$ is the fraction of links that node $j$  have with nodes of rank $i<j$. How these $\linksRC_j$ links are distributed between the $i<j$ nodes is given by $\massFunct(i)/\sum_{m=1}^{j-1}\massFunct(m)$.  Notice that in the case $j=2$ then $\massFunct(i)/\sum_{m=1}^{j-1}\massFunct(m)=1$, which implies that if there is one or several links between node $j=2$ and node $i=1$, that is $\linksRC_j\ne 0$,  these links are always shared between these two nodes. 
\begin{figure}
\begin{center}
\includegraphics{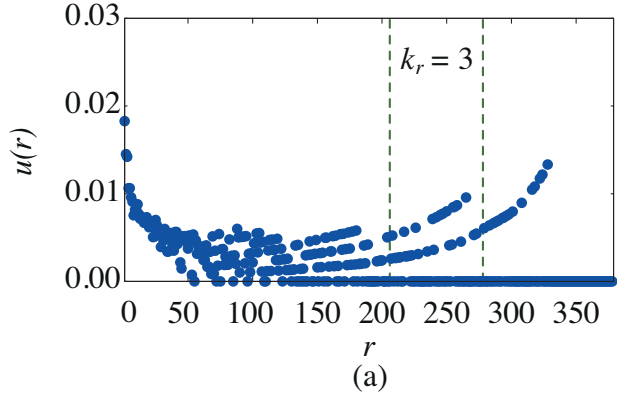}\\
\includegraphics{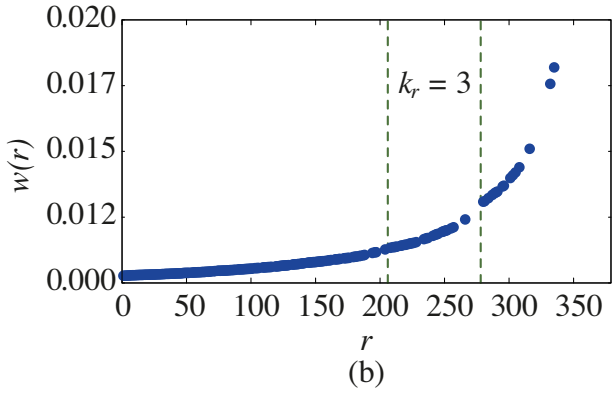}\\
\includegraphics{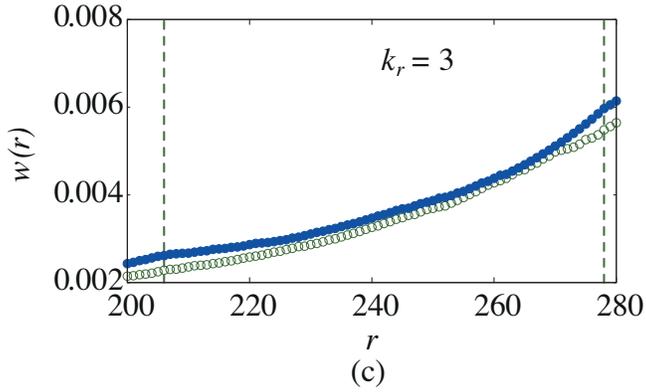}
\caption{\label{fig:maxEntNetSci}  Giant component of the co--authorship Network Scientists, the network has 379 nodes and 916 links. (a) The linking term $\massFunct(r)$ and (b) its rescaled version.  (c) Dependency of the MaxEnt solution with the ranking of the nodes.  For clarity we only plotted the solution of two different ranking schemes  for the case $k_r=3$, shown with an open and closed circle.}
\end{center}
\end{figure}

Figure~\ref{fig:maxEntNetSci}(a) shows a typical example of the function  $\massFunct(r)$, in this case the network is the giant component of the scientists working in the field of Complex Networks~\cite{Newman06}. The function $\massFunct(r)$ was obtained by solving numerically the MaxEnt problem. This network has characteristics that are interesting in our context. The nodes of degree one do not share any links with nodes of degree two. There is a tendency of nodes of similar low degree to connect with each other. This is expected as by construction all the co--authors of a paper are represented by a clique, that is, if there is an article with four authors, the degree of these nodes is as least four.  For small $r$, $\massFunct(r)$ captures the preferential attachment between nodes of high degree and low degree.  As the value of $r$ increases  $\massFunct(r)$ decreases and around $r>40$ increases again and has a 
{ 
discontinuous
}
shape. For large values of $r$, $\massFunct(r)$ increases as there is a preferential attachment between the low ranking nodes, capturing the property that there are papers with a small number of authors who form cliques. To describe the characteristics of $\massFunct(r)$ the first observation is that if $k_r$ is the degree of node $r$ and $\linksRC_r$ is the number of  links that connect to nodes of higher rank, then $k_r-\linksRC_r$ is the number of links that connect to nodes of lower rank. If $k_r-\linksRC_r=0$ means that a node $r'>r$ does not share a link with $r$. In other words,  $\massFunct(r)=0$ if $k_r-\linksRC_r=0$ which are the zeros shown in Fig.~\ref{fig:maxEntNetSci}(a). 
The 
{ discontinous}
 nature of $\massFunct(r)$ can be explained using the case $k_r=3$ with $k_r-\linksRC_r= i$ and $i>0$, (Fig.~\ref{fig:maxEntNetSci}(b)). If $i=1$ means 
 that from the three links that node $r$ has, only one link connects to nodes with rank $r'>r$, we denote the probability of this happening as $p$. Now if $i=2$, there are two links that can connect nodes $r'$ and $r$. If the MaxEnt solution is non--biased  then the probability that  node $r$ connects with $r'$ is $p+p$, that is the probability that one of the free links connects the two nodes plus the probability that the other free link connects the two nodes. In Fig.~\ref{fig:maxEntNetSci}(a), when $k_r=3$, the case $k_r-\linksRC_r=1$ corresponds to the lower 
 { ``branch''}
  and $k_r-\linksRC_r=2$ to the upper 
  { branch.
  }
    The implication of this observation is that the function $\stepFunct(r) = \massFunct(r)/(k_r-\linksRC_r)$  where $k_r-\linksRC_r \ne 0$ lies on a smooth curve.

{ 
From these observations Eq.~(\ref{eq:massResult}) can be rewritten as 
}
\begin{equation}
\label{eq:mainResult}
\eProb_{g(i,j)} = 
\frac{\stepFunct(i)\left(k_i-\linksRC_i \right) }{\sum_{n=1}^{j-1} \stepFunct(n)\left(k_n-\linksRC_n \right)}\frac{ \linksRC_j}{L},\quad i<j\\
\end{equation}
with $\eProb_{g(i,i)}=0$ and $\eProb_{g(j,i)} = \eProb_{g(i,j)}$ if $j<i$. The values of  $\stepFunct(i)$ are obtained recursively by using Eq.~(\ref{eq:mainResult}) in Eq.~(\ref{eq:constraintTwo}), that is
{
\begin{equation}
\label{eq:useConsTwo}
\frac{k_r}{L}+\sum_{j=r+1}^N \left( \frac{\stepFunct(r)\left(k_r-\linksRC_r \right) }{\sum_{n=1}^{j-1} \stepFunct(n)\left(k_n-\linksRC_n \right)}\frac{ \linksRC_j}{L}\right) = \frac{\linksRC_r}{L}
\end{equation}
}
which gives that for all $r$ 
\begin{equation}
\sum_{j=r+1}^N \frac{\linksRC_j}{\sum_{i=1}^{j-1}\stepFunct(i)(\degree_i-\linksRC_i)} = \frac{1}{\stepFunct(r)}.
\end{equation}
Taking the difference of two consecutive terms, i.e. $r=m$ and $r=m-1$ 
{ 
\begin{eqnarray}
\nonumber
\frac{1}{\stepFunct(m)}-\frac{1}{\stepFunct(m-1)}=\sum_{j=m+1}^N \frac{\linksRC_j}{\sum_{i=1}^{j-1}\stepFunct(i)(\degree_i-\linksRC_i)}-\sum_{j=m}^N \frac{\linksRC_j}{\sum_{i=1}^{j-1}\stepFunct(i)(\degree_i-\linksRC_i)}\\
=\frac{\linksRC_m}{\sum_{i=1}^{m}\stepFunct(i)(\degree_i-\linksRC_i)}=\frac{\linksRC_m}{\stepFunct(m)(\degree_m-\linksRC_m)+\sum_{i=1}^{m-1}\stepFunct(i)(\degree_i-\linksRC_i)}
\end{eqnarray}
and solving the above equation for $\stepFunct(m)$ we obtain
}
\begin{equation}
\label{eq:stepFunct}
\stepFunct(m) = \frac{\stepFunct(m-1)\sum_{i=1}^{m-1}\stepFunct(i)(\degree_i-\linksRC_i)}{\sum_{i=1}^{m-1}\stepFunct(i)(\degree_i-\linksRC_i)-\linksRC_m\stepFunct(m-1)}
\end{equation}
where  $\stepFunct(1) = 1$ without loss of generality. 
Eq.~(\ref{eq:mainResult}) in combination with Eq.~(\ref{eq:stepFunct}) define the probability that there is an interaction between two nodes of the null--model. 

{
Note that the constraints given by Eqs.~(\ref{eq:constraintOne})--(\ref{eq:constraintTwo}) were considered in Eq.~(\ref{eq:useConstOne}) and Eq.~(\ref{eq:useConsTwo}), respectively. The constraint Eq.~(\ref{eq:norma}) is satisfied as we formulated the problem using the normalised sequences. The Lagrangian multipliers $\lambda_{N-2+m}$ with $m=1,\ldots, N-2$ can be evaluated recursively from  Eq.~(\ref{eq:stepFunct}) using $w(m)=u(m)/(k_m-\linksRC_m)$ and $u(m) = \exp(\lambda_{N-2+m})$.

}

There is an ambiguity when labelling the nodes via a degree--dependent rank. For high degree nodes this is not a problem, as the degree tends to be unique so the rank labels these nodes unambiguously. For lower degree nodes, there are many nodes with the same degree. In this case the labelling of the nodes is not unique. 
We evaluated the MaxEnt solution using different ranking schemes for the nodes with equal degree. 
Fig.~\ref{fig:maxEntNetSci}(c) shows an example of how this ambiguity is reflected in the evaluation of the sequence $\stepFunct(r)$ obtained using 
 a random ranking scheme.
This ambiguity is reflected in the evaluation of the entropy and therefore in the null--model. We measured the change of entropy due to the re--labelling in different real networks and observed that the variation is very small (see Table~\ref{tab:one}) and has a minor effect on the properties of the null--model.

{
The table also show the effect  that the conservation of degree sequence and rich--club connectivity have in the properties of the null--model. 
The entropy of the \emph{C. elegans} and Power grid is larger than the entropy of the Internet, even that the number of nodes of these networks is smaller than the Internet.
Small entropy means that the number of networks that satisfy the constraints is also small, that is, the number of networks encompassed by the null model is small. The reason for this difference is that in the  AS--Internet, the top nodes tend to form a clique, a fully connected mesh~\cite{zhou2007structural}. This structure is not present in the \emph{C. elegans} or the Power grid network. The conservation of this clique imposes a strong restriction in the number of networks that have the same degree distribution and rich--club, this is reflected on the relatively small entropy per node in the AS--Internet. 
}

\begin{table}
\begin{center}
\begin{tabular}{l r r  c}\hline\hline
Network & nodes & links & entropy \\\hline
Karate club~\cite{zachary1977information} & 34 & 78 & $5.44 \pm 0.02$\\ 
\emph{C.~elegans}~\cite{Watts98}  & 297 & 2,148 & $34.06 \pm 0.02$\\
Power grid~\cite{Watts98} & 4,941 & 6,594 & $18.09\pm0.01$ \\
Internet & 22,963 & 48,436 & $17.734\pm 0.001$\\ \hline\hline
\end{tabular}
\caption{\label{tab:one} Variation of the entropy per node $\entrop/N$ due the ambiguity of the ranking scheme. The evaluation was done by randomising the ranks of nodes with equal degree. The averages were evaluated considering 40 randomised networks.
Internet data available from University of Oregon Route Views Project http://www.routeviews.org, dataset collected May 26, 2001.}

\end{center}
\end{table}

\section{Results}
\begin{figure}
\begin{center}
\includegraphics{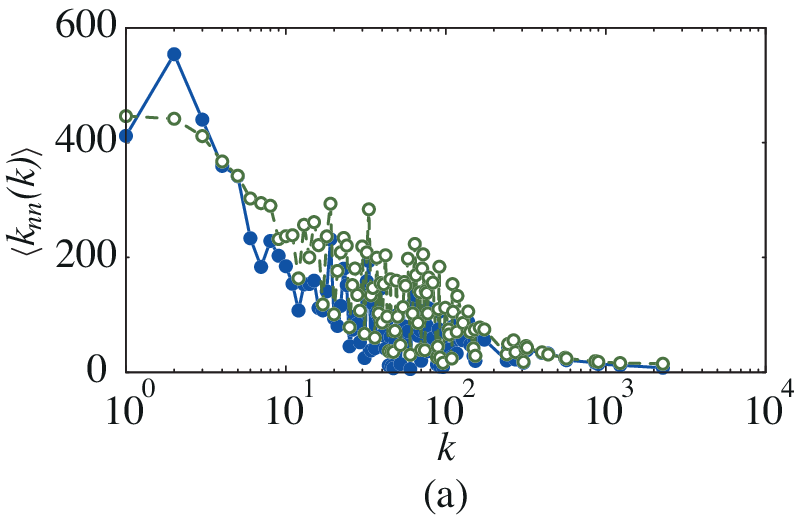}\\
\includegraphics{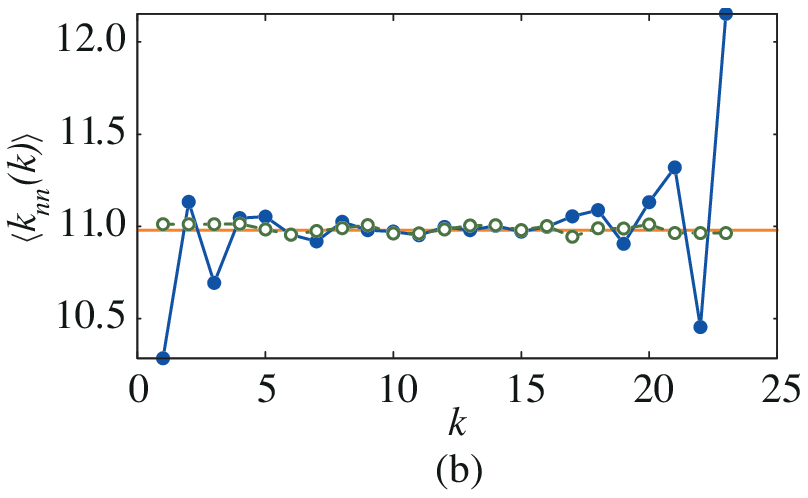}\\
\includegraphics{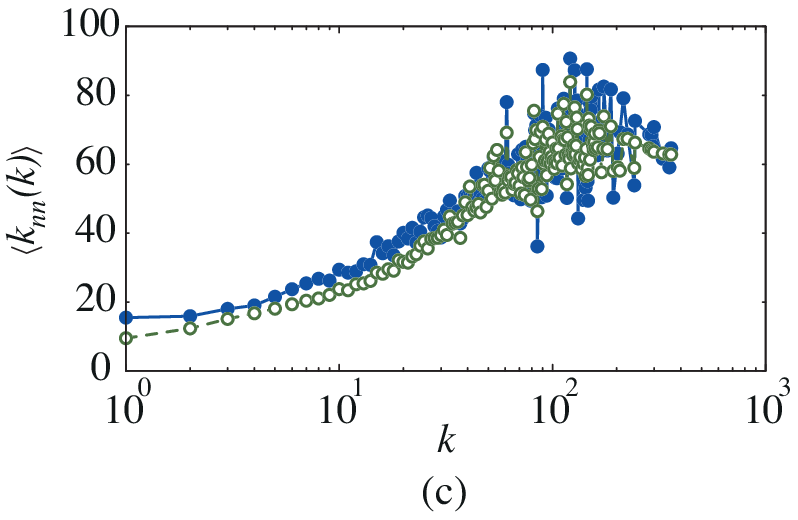}\\
\caption{\label{fig:wordAsso} Comparison of the average degree of nearest neighbours obtained from the data (filled squares) against the one obtained from the null--model (open circles). (a) For the protein (disassortative)  (b) for the random network (neutral)  and (c) for the astrophysics co--authorship (assortative)~\cite{Newman01}.
For the random graph $\langle k_{\rm{nn}} \rangle = \langle k^2 \rangle / \langle k \rangle$ (line) which is reproduced by the null--model. In this case the fluctuations seen in the data are due to statistical fluctuations in the evaluation of $\langle k_{\rm{nn}} \rangle$ from the data.}
\end{center}
\end{figure}

In order to evaluate how well the null--model can reproduce network properties that are related to the degree--degree correlation we compared the average degree of the nearest neighbours from the data against the null--model. 
For the null--model the evaluation of the expected degree is given by
\begin{equation}
\langle k_{\text{nn}}(k)\rangle = \frac{1}{N_k}\sum_{i=1}^{N}\left(\frac{1}{k}
\sum_{j=1}^{N} \eProb_{g(i,j)}L k_j\right)\delta_{k_i,k},
\end{equation}
where 
$\eProb_{g(i,j)}L$ is the number of links from node $i$  to node $j$ if the network has $L$ links.
In general we observed very good agreement for disassortative, neutral and assortative networks; see Fig~\ref{fig:wordAsso}. 

We also evaluated the number of links $n(k_i,k_j)$ that nodes with degree $k_i$ share with nodes of degree $k_j$. We compared the number of links obtained from the network data $n_D(k_1,k_j)$ against the number obtained from the null--model $n_N(k_1,k_2)$.
For the null--model the number of links is given by
\begin{equation}
n_N(k_i,k_j)=\sum_{m=1}^{N}\sum_{n=1}^{N} \eProb_{g(i,j)}L\, \delta (k_m,k_i)\delta (k_n, k_j)
\end{equation}
and their standard deviation is evaluated from the variance
\begin{equation}
var(k_i,k_j) = \sum_{m=1}^{N}\sum_{n=1}^{N}p_{g(m,n)}(1-p_{g(m,n)})L\, k_n  
\delta (k_m,k_i)\delta (k_n, k_j).
\end{equation}

\begin{figure}
\begin{center}
\includegraphics{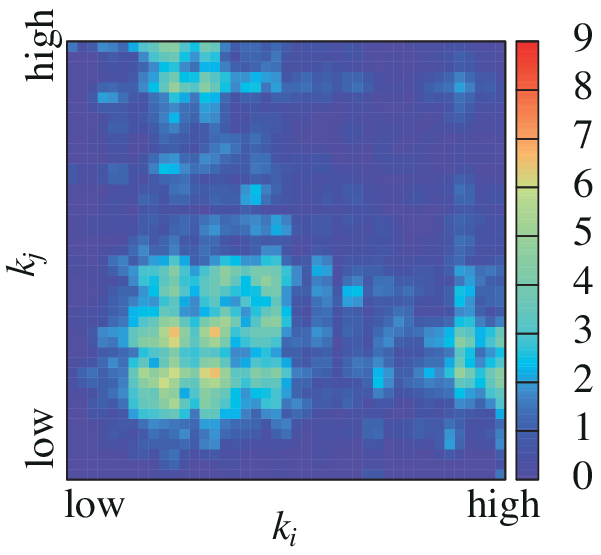}\\
\includegraphics{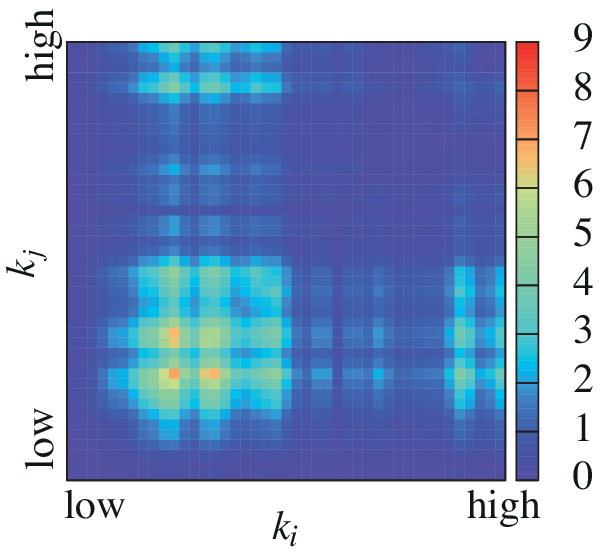}\\
\caption{\label{fig:lombriz} Number of links between nodes of degree $k_i$ and degree $k_j$ (top) for \emph{C.~elegans} network and (bottom) its  null--model. 
{
The null model captures the degree--degree correlation observed in the data.  In the figure, the connectivity of low degree nodes, both for  the null model and the data, show that these nodes tend to connect to nodes of similar degree or high degree nodes but not to other nodes.}
}
\end{center}
\end{figure}

 We compared these quantities, $n_D$ and $n_N$, for many different networks. Fig.~\ref{fig:lombriz} shows these quantities for the case of the \emph{C.~elegans}~\cite{Watts98}.  To check if the discrepancies between the two quantities are statistically significant we also evaluated the z--score as $z(k_i,k_j)=(n_D(k_1,k_j)-n_N(k_1,k_2))/\sigma_N(k_1,k_2)$ where $\sigma_N(k_1,k_2)$ is the standard deviation of the number of links obtained from the null--model. We noticed that the value of this score is low, less than two standard deviations, showing that for this case the null--model closely reproduces the degree--degree correlations of the data.

\section{Conclusions}
{
In summary, we have presented a 
technique
to construct null--models based on the maximal entropy  and the conservation of the degree sequence and rich--club coefficient.  As many real networks tend to have non trivial correlations between their nodes~\cite{Doro03}, 
the method 
presented here provides good null--models to study these networks, in particular 
scale--free networks where it is not possible to obtain a good approximation of the degree--degree correlation. 
For example, the null--models closely approximate
the correlations of assortative networks, which up to now has been a difficult property to reproduce with a null--model.
We envisage that the null--model presented here can be  useful when testing hypothesis where the rich--club plays a crucial part in the network's structure~\cite{Heuvel11,towlson2013rich}. 
From a practical context, the probability function describing the connectivity of the null--model was obtained without explicitly evaluating the Lagrangian multipliers. 
The method 
can be applied to generate null--models of large networks as the most onerous step is the node--ranking scheme which requires sorting in decreasing order the degree sequence. 
}

\section*{Acknowledgment}
Thanks to O.~Bandtlow, G.~Bianconi, V.~Latora, A.~Ma, A.~Mondrag\'on~B. and V.~Nicosia and the referees for useful comments and suggestions.


\end{document}